# Fully epitaxial fcc(111) magnetic tunnel junctions with a $Co_{90}Fe_{10}$/MgAlO/$Co_{90}Fe_{10}$ structure

Jieyuan Song,[1,2] Thomas Scheike,[1] Cong He,[1] Zhenchao Wen,[1] Tadakatsu Ohkubo,[1] Kazuhiro Hono,[1] Hiroaki Sukegawa,[1],* and Seiji Mitani[1,2]

[1] *National Institute for Materials Science (NIMS), Tsukuba, 305-0047, Japan*

[2] *Graduate School of Science and Technology, University of Tsukuba, Tsukuba, 305-8577, Japan*

*Email: sukegawa.hiroaki@nims.go.jp

**Abstract**

Magnetic tunnel junctions (MTJs) with bcc(001)-type structures such as Fe(001)/MgO(001)/Fe(001), have been widely used as the core of various spintronic devices such as magnetoresistive memories; however, the limited material selection of (001)-type MTJs hinders the further development of spintronic devices. Here, as an alternative to the (001)-type MTJs, an fcc(111)-type MTJ using a fully epitaxial CoFe/rock-salt MgAlO (MAO)/CoFe is explored to introduce close-packed lattice systems into MTJs. Using an atomically flat Ru(0001) epitaxial buffer layer, fcc(111) epitaxial growth of the CoFe/MAO/CoFe trilayer is achieved. Sharp CoFe(111)/MAO(111) interfaces are confirmed due to the introduction of periodic dislocations by forming a 5:6 in-plane lattice matching structure. The fabricated (111) MTJ exhibits a tunnel magnetoresistance ratio of 37% at room temperature (47% at 10 K). Symmetric differential conductance curves with respect to bias polarity are observed, indicating the achievement of nearly identical upper and lower MAO interface qualities. Despite the charge-uncompensated (111) orientation for a rock-salt-like MAO barrier, the achievement of flat, stable, and spin-polarized barrier interfaces opens a promising avenue for expanding the design of MTJ structures.

# 1. Introduction

A magnetic tunnel junction (MTJ) consisting of a ferromagnet (FM)/insulator (barrier)/FM trilayer is the fundamental structure of various spintronic devices such as read heads of hard disk drives, magnetic random-access memories (MRAMs), and highly sensitivity magnetic sensors.[1–5] In recent years, bcc FM(001)-based structures, such as Fe/MgO/Fe(001),[6,7] CoFe/MgO/CoFe,[8] CoFeB/MgO/CoFeB(001),[9,10] and Fe/MgAl$_2$O$_4$ (MAO)/Fe(001),[11–14] etc., have been widely used as an MTJ stack since large tunnel magnetoresistance (TMR) ratios at room temperature (RT) can be easily obtained due to the mechanism based on the specific bulk band structures, i.e., the spin-dependent coherent tunneling. This mechanism is based on the preferential tunneling of the $\Delta_1$ Bloch states through a (001)-oriented MgO or MAO barrier and the perfectly spin polarized $\Delta_1$ states of (001)-oriented bcc-FMs, such as Fe, Co-Fe, Co-based Heusler alloys.[15,16] This indicates that the (001)-type MTJ structure requires the stacking of different lattice systems, i.e., bcc-FM/fcc-type oxide barrier/bcc-FM structure, where the in-plane 45° rotation of the barrier lattice is required for the growth (Figure 1a).

To realize ultra-high-density MRAMs, a very large perpendicular magnetic anisotropy (PMA) is required to ensure long-term retention by achieving high thermal stability for MTJs with dimensions of less than 10 nm scale. In state-of-the-art spin-transfer-torque (STT) MRAMs using the (001)-type CoFeB/MgO/CoFeB MTJs, a combination of interface induced PMA at an ultrathin CoFeB/MgO(001) interface and bulk PMA of fcc(111)-based multilayers such as Co/Pt and Co/Pd has been employed.[17–19] However, the use of the different crystal systems for magnetic layers, i.e., 4-fold in-plane rotational symmetry of bcc(001)-based materials and 3- or 6-fold of fcc(111)-based materials, significantly limits the MTJ stack design and the available process temperature.

Recently, first-principles calculations predicted that a new class of an fcc(111)-type MTJ with Co/MgO/Co exhibits a large TMR ratio exceeding 2000%,[20] which is attributed to the interfacial resonance tunneling mechanism, in sharp contrast to the conventional coherent tunneling through bulk band structures (Figure 1b). A (111) plane generally has the most stable and lowest surface energy in a metallic fcc lattice,[21] allowing the construction of stable MTJ structures without changing the crystal system. In addition, the theoretical calculations of MTJs with Co-based $L1_1$(111) alloys, which have an fcc fundamental crystal structure, such as CoPt/MgO/CoPt(111) and CoPd/MgO/CoPd(111), exhibit both large TMR ratios and large PMA energies.[22] Therefore, fully fcc(111)-stacked MTJs will be promising to satisfy both a large TMR ratio and a large PMA energy for the scaling of spintronic devices. However, it is generally difficult to obtain a flat MgO(111) layer because of its very high surface energy due to the charge imbalance of its surface,[23] so it is necessary to establish preparation methods for fully fcc(111) based MTJ stacks.

In this article, we report the achievement of a fully epitaxial fcc(111) MTJ using $Co_{90}Fe_{10}$ (CoFe)/$Mg_4Al-O_x$ (MAO)/CoFe structure by combining magnetron sputtering for CoFe and electron-beam evaporation for MAO. Crystal structure analysis reveals epitaxial (111) growth with high crystallinity of the trilayer. Relatively flat interfaces were observed by cross-sectional scanning transmission electron microscopy (STEM) images with periodic misfit dislocations between the CoFe electrode and the MAO barrier to minimize the effect of their large lattice mismatch (~19%). We observed a TMR ratio of up to 37% at room temperature (RT) and 47% at 10 K in the CoFe/MAO/CoFe(111) MTJ. In addition, symmetric differential conductance ($dI/dV$) curves with bias polarities were observed, indicating the achievement of well-balanced interfaces between the lower and upper CoFe/MAO sides. The clear local structures are observed at around ±200 mV in the $dI/dV$ curve of the parallel (P) magnetization state, indicating the formation of specific interfacial states at the CoFe/MAO interfaces.

## 2. Results and discussion

Figure 2a shows the schematic MTJ stack structure with post-annealing temperatures and oxidation process conditions. The inset of figure 2b shows an atomic force microscopy (AFM) image for a 1×1 μm² area of the MTJ stack surface. We obtained an average roughness ($R_a$) 0.18 nm and the peak-to-valley (*P-V*) 1.96 nm from the AFM image, indicating the achievement of a flat surface suitable for an MTJ. Figure 2f-i show the reflection high energy electron diffraction (RHEED) patterns with electron-beam along the Al$_2$O$_3$ [10$\bar{1}$0] azimuth for (i) the Ru buffer layer, (h) bottom CoFe, (g) MAO barrier, and (f) top CoFe layers. All the patterns were taken after *in-situ* post-annealing for each layer. The patterns of the Ru buffer showed sharp streak lines with Kikuchi arcs. These features indicate epitaxial growth with a very flat surface and excellent crystallinity with hcp(0001) growth. The bottom CoFe layer also shows similar patterns, indicating that the epitaxial growth of fcc(111) CoFe is realized on the Ru(0001) plane. The MAO barrier also shows (111) epitaxial growth; however, the spotty pattern indicates that the MAO surface is rougher than the bottom CoFe surface as known in MgO(111).[24–26]. This may be attributed to the large lattice mismatch with CoFe and the occurrence of reconstruction to reduce the surface energy of the charge-uncompensated (111) surface. Nevertheless, the top CoFe layer recovers the (111) epitaxial growth with high crystallinity on the MAO barrier, demonstrating the achievement of a fully epitaxial fcc(111) MTJ.

Figure 2b shows the out-of-plane 2θ-ω XRD spectrum of the MTJ stack. The (0001) growth of hcp Ru and the (111) growth of fcc CoFe were confirmed by the Ru(0002), Ru(0004), CoFe(111), and CoFe(222) peaks. The MAO(111) and (222) peaks (indices for a cation-disordered spinel structure[12]) were also observed. Due to the achievement of the flat Ru buffer, fringe patterns corresponding to 40 nm are clearly observed around the Ru peaks. Figure 2c shows the in-plane XRD pole scan (φ-scan) spectra to obtain information on fractions of hcp and fcc components in the CoFe layers. The upper (lower) spectrum corresponds to the pole scan of (11$\bar{1}$)$_{fcc}$, $2\theta_\chi \sim 43.9°$ and $\chi \sim 19.9°$ [(1$\bar{1}$01)$_{hcp}$, $2\theta_\chi \sim 46.2°$ and χ

~ 29.9°].[27] The upper spectrum shows the distinct 6-fold peaks. The additional smaller 6-fold peaks at an offset of 30° are due to the presence of a variant in the CoFe layers. In contrast, the lower spectrum shows no distinct peaks. Thus, the CoFe layers of the MTJ stack consist of an almost perfect fcc structure using the $Co_{90}Fe_{10}$ composition instead of pure Co. Figures 2d and 2e show the rocking curves ($\omega$-scan) of the Ru(0002) peak and the CoFe(111) peak of a reference sample with a bottom electrode structure of $Al_2O_3$(0001) sub.//Ru (40 nm)/CoFe (20 nm)/Ru (2 nm). These curves have two components: a sharp specular component and a broad diffuse component. Such broad components in a rocking curve are often observed in highly oriented epitaxial thin films due to the strain field caused by introduced misfit dislocations near interface regions.[28] The full width half maximum (FWHM) values of the sharp and broad components for Ru (CoFe) are 0.054° and 0.34° (0.057° and 0.38°), respectively. The small FWHMs of the sharp components indicate the achievement of nearly perfect orientation for Ru(0001) and CoFe(111).

HAADF-STEM images of the MTJ cross-section are shown in figure 3. From the HAADF-STEM image of the entire stack (figure 3a), flat and sharp interfaces were maintained from the Ru buffer to the top CoFe layer. Figure 3b shows the magnified image across the CoFe/MAO/CoFe trilayer. The lower CoFe/MAO interface is atomically flat. The upper MAO/CoFe interface is slightly rougher than the lower interface; however, the formation of the stable fcc-based (111) barrier layer shown in the image is the important step in the design of fully (111)-type MTJs. The MAO thickness was determined to be ~2.75 nm from the HAADF-STEM image in figure 3b. Figure 3c shows the EDS elemental line profiles across the cross-section of the film. Note that the O signal outside the barrier is the artifact coming from the surface of the TEM specimen. The MAO barrier consists mainly of MgO with a negligible amount of Al, which can be attributed to the low Al concentration of the electron-beam (EB) source ($Mg_4Al-O_x$) and the change in Mg-Al composition during the deposition process. A small amount of Fe segregation was also observed at the upper and lower CoFe/MAO interfaces, which could be due to the higher oxygen affinity of Fe than Co. The NBED patterns taken from the region of the HAADF-STEM observation (figure 3d-f) confirm the epitaxial growth of (111) orientation of the bottom CoFe, MAO barrier, and top CoFe layers. The epitaxial relationship of $Al_2O_3(0001)[11\bar{2}0] \parallel Ru(0001)[10\bar{1}0] \parallel Co_{90}Fe_{10}(111)[11\bar{2}] \parallel MAO(111)[11\bar{2}]$ was determined by the NBED patterns, which is consistent with the RHEED and XRD results [see the model of figure 1b]. Thus, our CoFe/MAO/CoFe(111) structure almost reproduced the theoretical supercell stack of Co/MgO/Co(111), except for the large lattice mismatch, as explained next.

Figure 4 shows the high-magnification HAADF-STEM image near the barrier region. According to the local atomic structure of MAO, its lattice parameter is determined to be $a_{MAO}$ = 0.420 nm, which is almost the same as the bulk value of MgO (0.421 nm). The lattice mismatch between CoFe and MAO is calculated to be ~19%. Due to the large lattice mismatch, a periodic distribution of dislocations at the lower MAO/CoFe interface is revealed in the inverse fast Fourier transformation (FFT) image in figure 4b, where six {111} planes of CoFe coincide with five {111} lattice planes of MAO. This structure is illustrated in figure 4c. The distance of the 5-plane MAO domain and that of the 6-plane CoFe is 1.032 nm and 1.086 nm, respectively; therefore, their mismatch becomes only –5% by introducing the 5:6 domain matching, which is much smaller than the 19%. The top CoFe was not perfectly oriented and may have rotated slightly along the [111] axis within the film plane, as evidenced by the interplanar distance in the top CoFe layer. It is also suggested that at the interfaces the O atoms are located directly above the (Co, Fe) atoms, which is consistent with the theoretical calculation predicting an energetically stable Co-O interface rather than a Co-Mg interface.[20] Therefore, it is speculated that the (111) growth was maintained from the bottom to the top CoFe layer. The observed 5:6 domain matching due to the introduction of the periodic dislocations may be one of the reasons for the formation of the relatively flat barrier interfaces.

Figure 5a and b show the magnetotransport four-probe measurement setup of an MTJ pillar. Figure 5c shows the TMR ratios as a function of in-plane magnetic field $H$ along $Al_2O_3[10\bar{1}0]$ at 300 K (RT) and 10 K (bias voltage <10 mV). Exchange spin-valve type loops were obtained due to the use of a synthetic antiferromagnetic (SAF) structure (i.e., $Co_{90}Fe_{10}$/Ru/$Co_{50}Fe_{50}$/IrMn, see figure 2a) in the MTJ stack. TMR ratios of 37% at RT and 47% at 10 K were observed from the MTJ. A resistance area product ($RA$) for the P state at RT was $\sim 2.7 \times 10^6$ $\Omega \cdot \mu m^2$, which is larger than the value of the recent Fe/MgO (2.75 nm)/Fe(001) MTJ ($\sim 6 \times 10^5$ $\Omega \cdot \mu m^2$),[7] which may indicate the difference in the transport mechanism from the $\Delta_1$ coherent tunneling. However, the observed TMR ratios are still much smaller than the theoretical value in Co/MgO/Co(111) (~2000%). The formation of the interfacial resonance states, which is responsible for the large theoretical TMR ratio, can be significantly suppressed when imperfections at FM/barrier(111) interfaces for both sides are observed, e.g., misfit dislocations, atomic diffusion, and roughness, etc. Our STEM observations revealed the introduction of many interfacial dislocations and atomic steps at the lower and upper CoFe/MAO interfaces due to the inevitable lattice mismatch, which may significantly reduce the TMR ratio due to the absence of the interfacial resonance effect in our MTJ.

Figure 5e (figure 5f) shows the temperature dependence of the TMR ratio ($R_{AP}$ and $R_P$, where $R_{AP}$ is the resistance of the antiparallel (AP) magnetization state and $R_P$ is the resistance of the parallel (P) magnetization state). A monotonic increase in the TMR ratio with decreasing temperature was observed. Similarly, both $R_{AP}$ and $R_P$ show a monotonic increase with decreasing temperature. $R_{AP}$ shows a stronger change than $R_P$; therefore, the temperature dependence of the TMR ratio is mainly determined by the dependence of $R_{AP}$. Such a feature is also observed in the conventional bcc(001) MTJs.[7,13] Note that the slight increase in $R_P$ with decreasing temperature is in contrast to the highly (001)-oriented Fe/MgO/Fe and Fe/MAO/Fe MTJs, which show a slight decrease in $R_P$.[7,14]

Figure 5d and 5g-h show the bias voltage $V$ dependences of (d) TMR ratio (TMR-$V$ curve), (g) and (h) differential conductance of the AP state ($G_{AP}$) and P state ($G_P$) at RT and 10 K. The $G_{AP}$ and $G_P$ were obtained by calculating the derivatives of each current-voltage curve (i.e., $dI/dV$). The symmetric curves with respect to the bias polarity were observed for all the curves, indicating the achievement of almost the same quality of the lower and upper CoFe/MAO interfaces, i.e., similar interfacial electronic states. The TMR-$V$ curves show a monotonic decrease with increasing $|V|$, similar to conventional MTJs. The $V_{half}$ values, $V$ at which a TMR ratio becomes half of the zero bias value, were estimated to be 0.49 and −0.51 V (0.41 and −0.46 V) at RT (10 K) for the positive and negative bias polarities, respectively. No distinct structures were observed in the TMR-$V$ curves even at 10 K. In contrast, the conductance curves show some fine local structures. Parabolic-like curves with a dip structure near zero bias are observed in the $G_{AP}$ curves, where the dip is more pronounced at 10 K. The zero bias dip in the AP state may be due to magnon-assisted inelastic tunneling, which is also commonly observed in various MTJs.[29–31] In the $G_P$ curve at RT, a broad (bias independent) plateau is observed in the $V$ range between −0.15 and 0.15 V as indicated by an open square bracket in figure 5h. At 10 K, a small dip appears near zero bias and two local minimum structures become pronounced at $V$ ~0.28 and ~−0.26 V, as indicated by arrows. Similar local minimum structures in the $G_P$ curve have been reported in bcc(001) CoFeB-based MTJs at around $|V|$ = 0.23~0.35 V.[32,33] It is currently unknown whether the origin of these local minimum structures in the $G_P$ curves is identical in the bcc(001) MTJ and the fcc(111) MTJ. It is expected that more detailed information on the interface states will be obtained when fcc(111) MTJs with reduced interface imperfection are realized. In general, the local structures in the $G_P$ curve reflect tunneling processes through electronic states near the electrode/barrier interfaces.[34] Therefore, the transport process in the fcc(111) MTJ structures, which exhibit significant interfacial resonance tunneling should be different from that in the bcc(001) MTJ structures. It is also expected that the reduction of the interfacial dislocation density can be a key to improve the TMR ratio of the fcc(111) MTJs by enhancing the interfacial resonance tunneling. The use of lattice-matched systems by tuning the MAO composition[11,35] and by using FM materials with larger lattice constants, such as CoPt and CoPd, will be a promising way to fabricate high-quality fcc(111) MTJs.[22]

3. Conclusion

We obtained a fully (111) epitaxial MTJ in CoFe/MAO/CoFe structure using a combination of magnetron sputtering and EB evaporation. The high crystallinity with epitaxial MAO(111) barrier was obtained even though there is a large lattice mismatch between the CoFe and MAO. The relatively flat MAO interfaces were achieved by introducing periodic misfit dislocations, which resulted in a peculiar 5:6 lattice matching. TMR ratios of 37% at RT and 47% at 10 K were observed from the CoFe/MAO/CoFe(111) MTJ. The differential conductance curves of the MTJ were symmetric with respect to bias polarity, indicating the well-balanced interfaces of the lower and upper CoFe/MAO sides. The observed local structures in differential conductance of the P state may reflect a specific interface structure of the fcc(111) interfaces; however, the small TMR ratios suggest our MTJ does not show significant TMR enhancement due to the interfacial resonance tunneling mechanism that the theory predicted. Nevertheless, our TMR demonstration by achievement of a fully epitaxial CoFe/MAO/CoFe(111) structure can accelerate the development of fcc(111)-based MTJs, which is useful for a future high-density MRAM and highly sensitivity magnetic sensor applications.

4. Experimental Section

MTJ multilayers were deposited on a single-crystal sapphire $Al_2O_3$(0001) substrate using a magnetron sputtering apparatus (ULVAC, Inc.) with a base pressure of ~4 × $10^{-7}$ Pa combined with an EB evaporator for a Mg$_4$Al-O$_x$ barrier.[14] Top-exchange-bias type MTJ multilayers were prepared with the following stack design: Al$_2$O$_3$(0001)//Ru (40)/Co$_{90}$Fe$_{10}$ (CoFe) (20)/Mg (0.5)/Mg$_4$Al-O$_x$ (2.5)/CoFe (5)/Ru (0.75)/Co$_{50}$Fe$_{50}$ (2.2)/Ir$_{20}$Mn$_{80}$ (IrMn) (10)/Ru (10) (nominal thickness in nm) (Figure 2a). Instead of pure Co assumed in the theory,[20] which shows both hcp and fcc structures, the Co$_{90}$Fe$_{10}$ layers were used as FM layers to obtain a single fcc phase. The Mg-rich MAO (Mg$_4$Al-O$_x$) was used as a barrier in this study since larger TMR ratios (up to 429% at RT)[14] than those of MgO (417%)[7] were observed in the previous Fe(001)-MTJ studies. Prior to deposition, the sapphire substrates were *ex-situ* annealed at 1000°C for 1 h in a muffle furnace under an air atmosphere to improve their surface flatness. All metallic layers were deposited by DC magnetron sputtering in which Ru is deposited at 350°C[36] and others were deposited at RT. Each layer was *in-situ* post-annealed to obtain a flat surface and good crystallinity. The Mg layer was inserted at the lower CoFe/MAO interface to protect against over-oxidation during the MAO deposition.[37] The MAO barrier was deposited by EB-evaporation from a sintered MAO block with a nominal Mg/Al atomic ratio = 4 at a deposition rate of ~8 × 10$^{-3}$ nm s$^{-1}$.[14] The CoFe/Ru/Co$_{50}$Fe$_{50}$/IrMn structure is a pinned layer with a SAF structure to obtain a stable AP state. After the deposition, the MTJ stack was *ex-situ* annealed at 300°C under a 7 kOe magnetic field along the Al$_2$O$_3$[10$\bar{1}$0] direction. The crystal structures were characterized by *in-situ* reflection high energy electron diffraction (RHEED) and *ex-situ* X-ray diffraction (XRD) with Cu $K_\alpha$ radiation (wavelength: 0.15418 nm) using a graphite monochromator. The microstructure analysis was characterized by the high-resolution high-angle annular dark-field STEM (HAADF-STEM), nano-beam electron diffraction (NBED), and the energy dispersive x-ray spectroscopy (EDS) (FEI Titan G2 80–200 ChemiSTEM). The MTJ stacks were patterned by photolithography and Ar-ion milling into 39 μm$^2$ area elliptical junctions with the long axis parallel to the Al$_2$O$_3$[10$\bar{1}$0] axis. TMR ratios were measured by a conventional DC 4-probe method at RT with a source meter (Keithley, 2400) and nanovoltmeter (Keithley, 2182A). Temperature dependence from RT to 10 K of the TMR ratio and resistances were characterized using a physical property measurement system (PPMS) (Quantum Design, Dynacool). In this study, a negative bias voltage corresponds to electrons tunneling from the bottom layer to the top layer. The TMR ratio is defined as $(R_{AP} - R_P)/R_P \times 100\%$.

## Acknowledgements

We are grateful to Keisuke Masuda and Yoshio Miura for their fruitful discussion on the transport mechanism of (111)-MTJs, to Jun Uzuhashi and Kyoko Suzuki for their assistance in the preparation of the TEM samples, and to Hiromi Ikeda for her support in the MTJ microfabrication. This work was partly supported by JST CREST (JPMJCR19J4), JSPS KAKENHI Nos. 21H01750 and 22H04966, and Samsung Advanced Institute of Technology. J.S. acknowledges the National Institute for Materials Science for the provision of a NIMS Junior Research Assistantship.

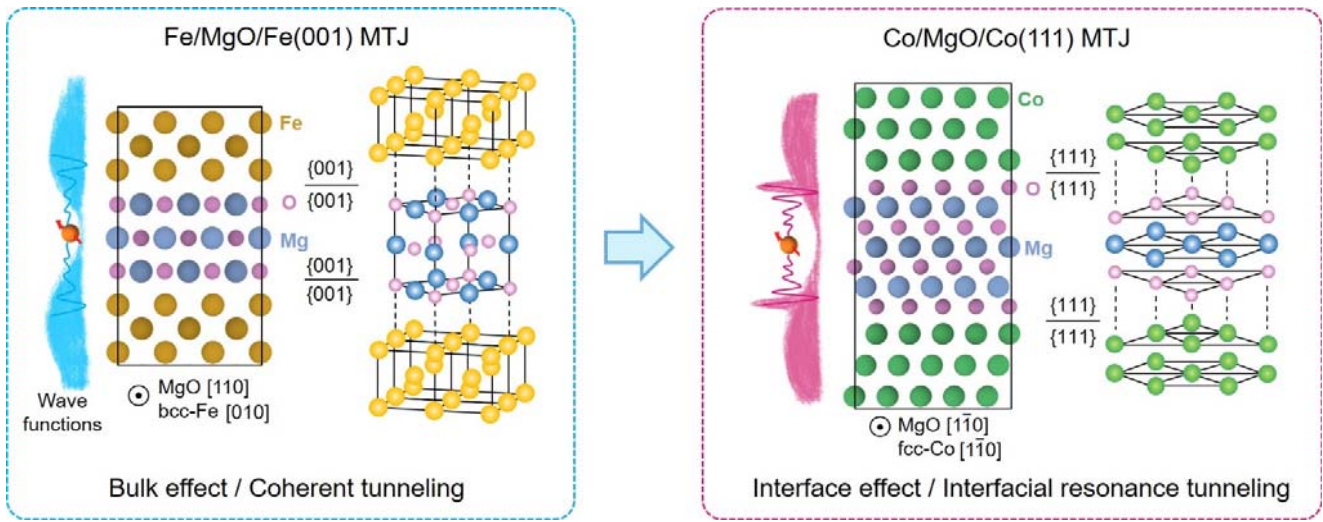

**Figure 1.** Concept of fcc(111) MTJ. Illustration of crystallographic relationships of conventional bcc Fe/MgO/Fe(001) (left) and fcc Co/MgO/Co(111) MTJ (right). The theoretical TMR mechanism of the Fe/MgO/Fe(001) is based on *bulk coherent tunneling*, while that of the Co/MgO/Co(111) is based on *interfacial resonance tunneling*.

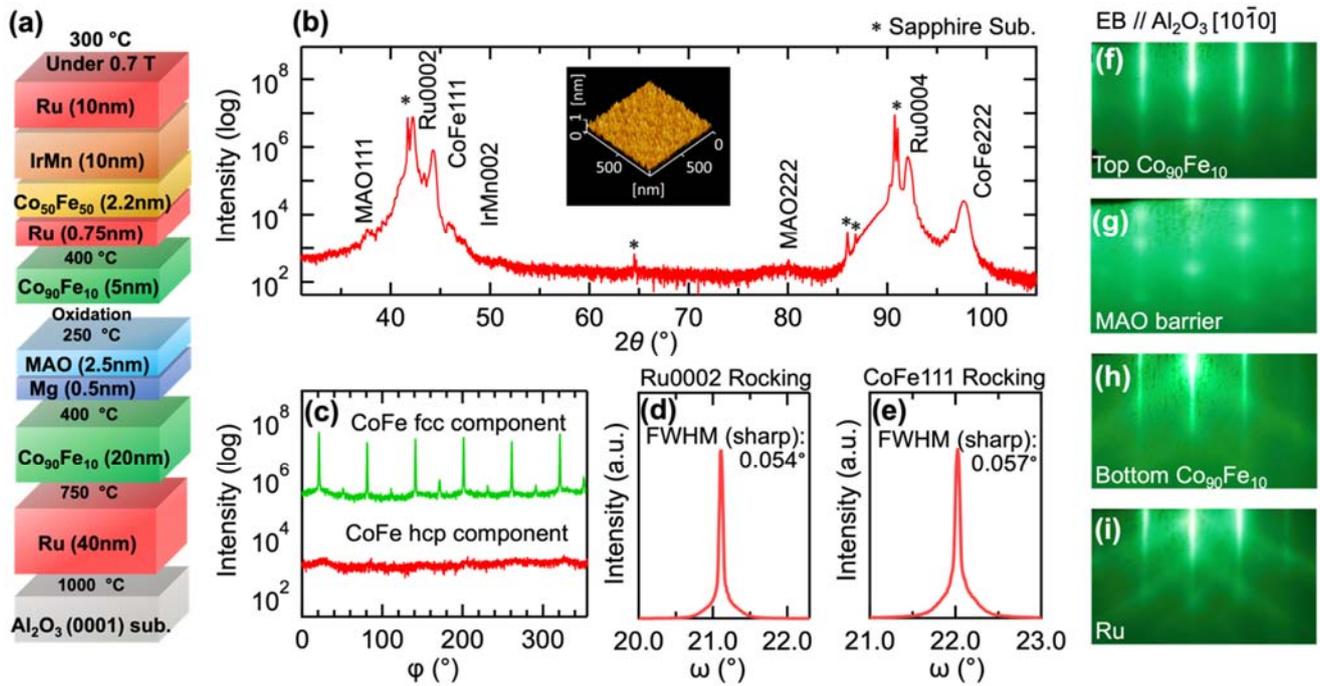

**Figure 2.** (a) Schematic MTJ stacking structure with post-annealing temperatures and oxidation process. (b) Out-of-plane XRD scans of an MTJ stack. (c) In-plane pole figures for CoFe($11\bar{1}$)$_{fcc}$ and ($1\bar{1}01$)$_{hcp}$ poles of the MTJ stack. (d, e) Rocking curves of (d) Ru(0002) peak and (e) CoFe(111) peak of a stack with $Al_2O_3$(0001)//Ru (40 nm)/CoFe (20 nm)/Ru (2 nm). (f-i) RHEED patterns of (f) top CoFe, (g) MAO barrier, (h) bottom CoFe, and (i) Ru buffer. The incident electron beam is parallel to the $Al_2O_3$[$10\bar{1}0$] azimuth. Inset of (b): AFM image of the MTJ stack. All observations were done after post-annealing.

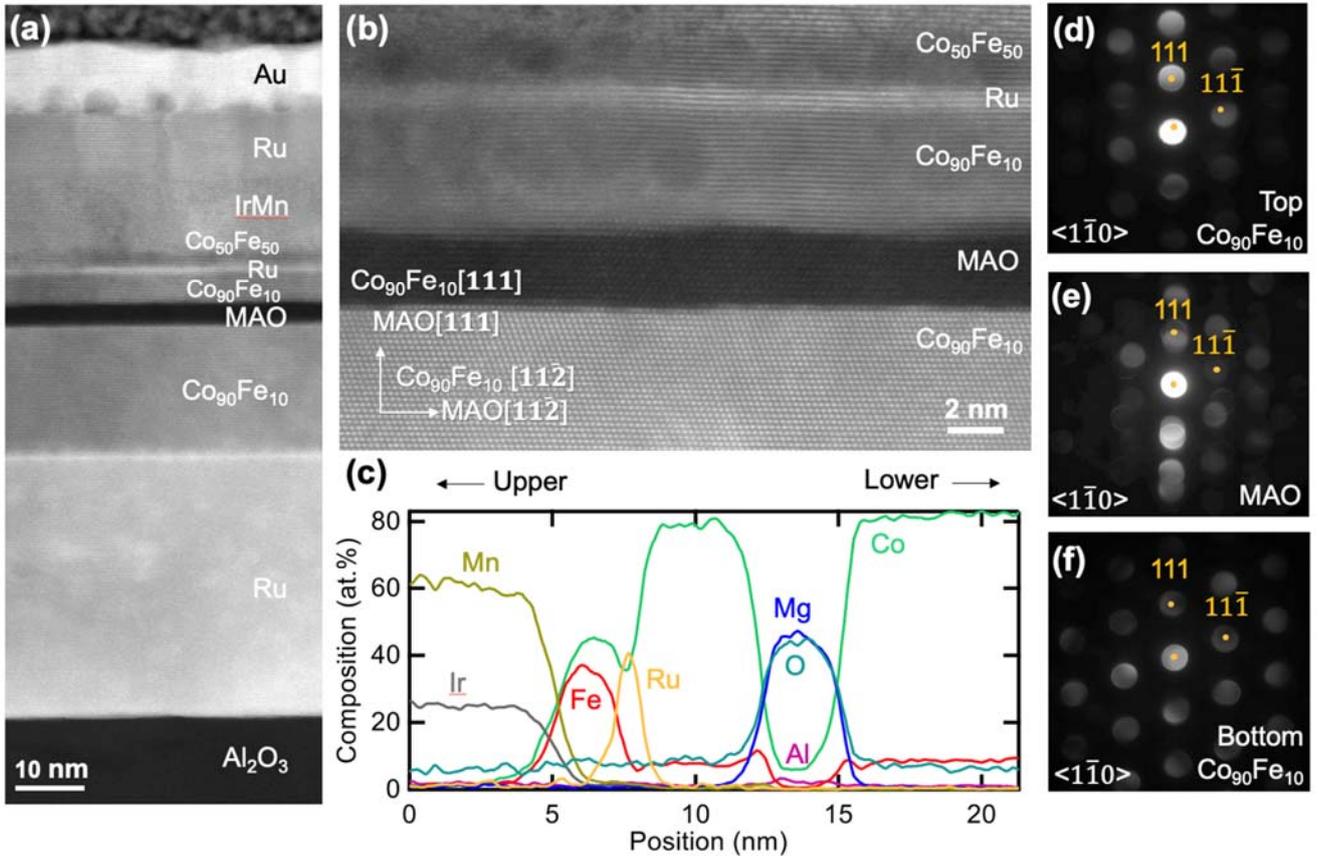

**Figure 3.** (a) Cross-sectional HAADF-STEM image of the MTJ stack observed along $Al_2O_3[10\bar{1}0]$. (b) Magnified image of (a) near the CoFe/MAO/CoFe. (c) Elemental depth profiles using EDS. (d-f) NBED patterns for (d) top CoFe, (e) MAO barrier, and (f) bottom CoFe.

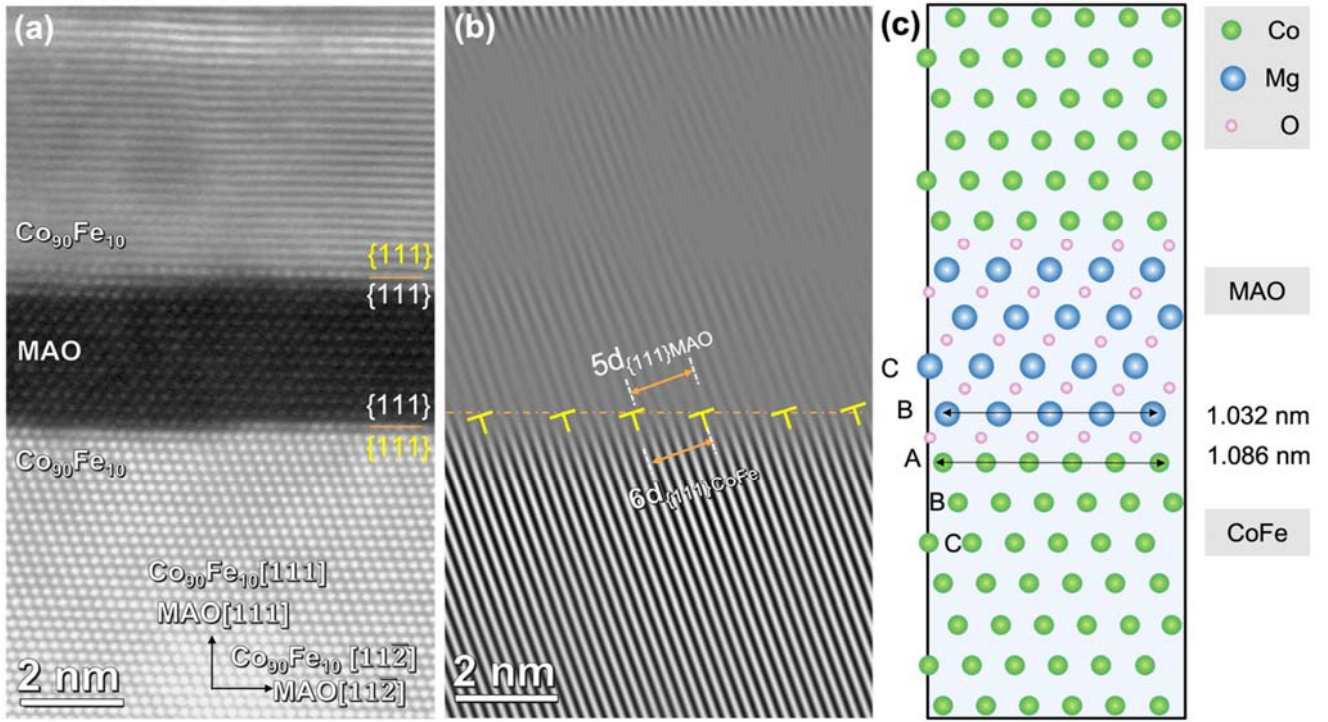

**Figure 4.** (a) Cross-sectional HAADF-STEM images of the MTJ stack observed along $Al_2O_3[10\bar{1}0]$, orange lines indicate the {111} planes of CoFe and MAO. (b) FFT filter images using (a). ⊥ Marks in (b) indicate the lattice dislocations at the interface. (c) Illustration of the interface atomic model.

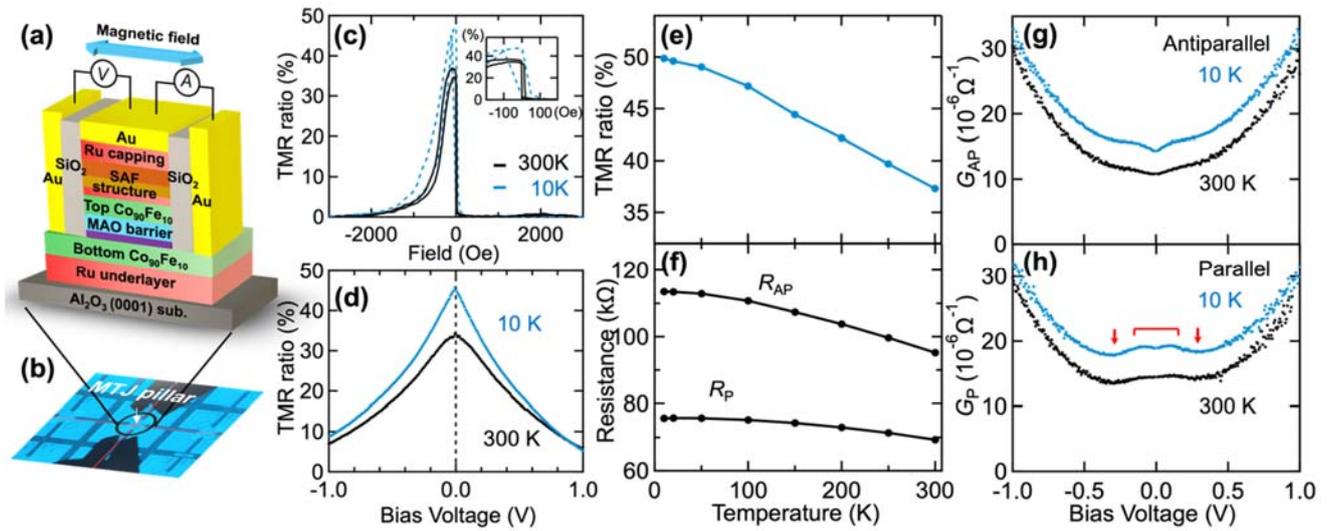

**Figure 5.** (a) Schematic of a patterned MTJ with the stack structure with the four-probe measurement setup. (b) Photo of the four-probe measurement. (c) TMR-$H$ curves at 300 K (solid, black) and 10 K (dashed, blue) for the MTJ. (d) Bias voltage dependence of TMR ratio at 300 K and 10 K. (e) and (f) Temperature dependence of (e) TMR ratio and (f) $R_P$ and $R_{AP}$. (g) and (h) $G$ spectra at 300 K and 10 K for (g) AP and (h) P states. For (g) and (h), the 10 K curves are shifted upward (by $4 \times 10^{-6}$ $\Omega^{-1}$) for comparison. The arrows and open square brackets in (h) are the local structures. Inset of (c): TMR-$H$ curves at a low field region.